\documentclass{ws-ijgmmp}
\def\be{\begin{equation}}
\def\ee{\end{equation}}
\def\ba{\begin{eqnarray}}
\def\ea{\end{eqnarray}}

\usepackage{slashed}

\usepackage{epsfig}
\def\be{\begin{equation}}
\def\ee{\end{equation}}
\def\bea{\begin{eqnarray}}
\def\eea{\end{eqnarray}}

\def\yzero{\smash{\hbox{$y\kern-4pt\raise1pt\hbox{${}^\circ$}$}}}

\def\beq{\begin{equation}}
\def\eeq{\end{equation}}
\def\beqa{\begin{eqnarray}}
\def\eeqa{\end{eqnarray}}

\def\-{\hphantom{-}}

\def\s2{\frac{1}{\sqrt2}}

\def\beq{\begin{equation}}
\def\eeq{\end{equation}}
\def\beqa{\begin{eqnarray}}
\def\eeqa{\end{eqnarray}}

\def\IF{\relax{\rm I\kern-.18em F}}
\def\II{\relax{\rm I\kern-.18em I}}
\def\IP{\relax{\rm I\kern-.18em P}}
\def\IC{\relax\hbox{\kern.25em$\inbar\kern-.3em{\rm C}$}}
\def\IR{\relax{\rm I\kern-.18em R}}

\def\Dsl{\,\raise.15ex\hbox{/}\mkern-13.5mu D} 
\def\IZ{Z\kern-.4em  Z}

\begin{document}

\markboth{M.P. Garcia del Moral}
{Geometry of Gauging Processes}

%
\catchline{}{}{}{}{}

\title{ On the Geometry of Sculpting-like Gauging Processes}
\author{Mar\'\i a Pilar Garc\'\i a del Moral}
\address{ Depto. de F\'\i sica, Universidad de Oviedo, Avda Calvo
Sotelo S/n. Oviedo,33007 Spain and Instituto de F\'\i sica Te\'orica IFT-UAM/CSIC, Cantoblanco, 28049 Madrid, Spain\\
\email{garciamormaria@uniovi.es} }
\maketitle
\begin{history}
\received{(Day Month Year)}
\revised{(Day Month Year)}
\end{history}

\begin{abstract}
 Recently, a new gauging procedure called Sculpting mechanism was proposed to obtain the M-theory origin of type II gauged Supergravity theories in 9D. We study this procedure in detail and give a better understanding of the different deformations and changes in fiber bundles, that are able to generate new relevant physical gauge symmetries in the theory. We discuss the geometry involved in the standard approach (Noether-like) and in the new Scultping-like one and comment on possible new applications.
\end{abstract}
\section{Introduction}
Symmetries play a fundamental role in Physics. They are coordinate/field transformations that leave invariant a functional defined in terms of these fields (action, Hamiltonian). The so-called  "Gauging a theory" is a physical deformation/change procedure in which an initial action characterizing a physical system is transformed into an inequivalent new one, with new gauge groups that include new interactions in the system. The initial un-deformed theory possesses some  global symmetries that are promoted into local ones in the final theory.  The new action contains these new local symmetries by means of a minimal coupling of the matter fields with a gauge field (connection). Derivatives are substituted by covariant derivatives carrying the interaction. This gauging procedure is usually done "locally" since the bundle where the connection is defined is assumed to be trivial. The final action must be invariant under the new local symmetry (Noether Mechanism). The Sculpting gauging is an alternative proposal based on extracting a gauge symmetry by modifying the initial theory imposing some topological restrictions on the action. It is formulated in terms of bundles.
 In this paper we extend the analysis in \cite{gm} and \cite{gmpr}, and provide two new examples (Section 3, examples one and two). In Section 3 example three, we extend previous analysis \cite{gm}, and explain each step in the process of topological change and deformation of the supermembrane bundle. Section 4, is devoted to conclusions. In addition we give general remarks about the scheme to follow to implement this sculpting procedure to the cases of a D4-brane worldvolume action.
\section{Standard Gauging Procedure: The Noether Mechanism} In a geometric context \cite{kn}, the fields of a physical Field Theory are often sections on vector bundles $E\to B$, associated to a principal G-bundle $P$. These construction however can be generalized to arbitrary fiber bundles -as we will see along the examples of Section 3, in which the sections act on manifold bundles-. In the initial formulation, the vector bundles are typically taken trivial. The action integral of a physical  theory is a functional $S$ defined on
sections $s$ acting on associated bundles $E$, $s\in\Gamma(E)$, which are invariant under certain global symmetries. These \emph{global} transformations  are defined by constant maps $a: B\to H$, where $B$ is the base manifold of dimension $m$ and H is a certain group representing the global symmetry group, subgroup of the General Linear Group of matrices, $GL(m,\mathbb{C})$ . Then, the sections $s$ transform under these global transformations into new ones as
$\widehat{s}=a s$, $\frac{\partial\widehat{s}}{\partial x_i}=\widehat{\left(\frac{\partial s}{\partial x_i}\right)}.$
Derivatives are always present in the invariant action $S$ -they always appear in the kinetic terms, and may also appear in the potential.- When the gauging process is performed, the global symmetry transformation $a$ is promoted into a \emph{local} one $a^{'}=a(x)$. 
  Thus, a gauge transformation of $E$ is given by a nonconstant smooth map $a^{'}:B\to{\rm GL}(m,\,{\mathbb C})$ and the former functional will not be invariant under this gauge transformation. Derivatives of the sections add an extra term $\frac{\partial a^{'}}{\partial x_i}s$ and to compensate it, one introduces a connection $\nabla$ in the vector bundle $E$ and substitutes the usual derivatives of sections with covariant derivatives in the expression of the action functional $
\partial_i\to\nabla_i:=\frac{\partial}{\partial x_i}\lrcorner \nabla.
$
 The invariance under a local symmetry of the theory implies gauge invariance,-let us denote by $T$- of the action functional. Let us denote the transformed connections on $E$ as the
operator $\widetilde\nabla$ defined on $s$ by $\widetilde\nabla(s)=T(\nabla(T^{-1}(s)).$ The new transformed sections under $T$ transformations as $\widetilde{s}$ satisfy,
$
\widetilde{s}=T(s)=T\circ s$, and $\widetilde{\nabla}\widetilde{s}=(T\circ\nabla\circ T^{-1})(T\circ s)=\widetilde{(\nabla s)}.
$\\
Let $f$ be a local frame  for $E$ on an open set $U\subset B$, then $T(f_a)=\sum_b f_b t^b_a$, where $t$ is a nonsingular
matrix of functions defined on $U$ . We denote by $\alpha$ the dual frame for $E^*$, so $T$ can be written as $T=f\cdot t\cdot \alpha$.
If $A$ is the connection form of $\nabla$ in the frame $f$, then
$$\widetilde\nabla f =T(\nabla(f t^{-1}))=T(f At^{-1}+f dt^{-1})=f t\alpha(f A t^{-1}+f dt^{-1})=f(tAt^{-1}+tdt^{-1}).$$
That is, the connection form of $\widetilde\nabla$ in $f$ is $tAt^{-1}+tdt^{-1}.$ By an iterative process necessary terms are added to recover the gauge invariance  of the  modified action functional.
\paragraph{A $U(1)$ example} Take for example the Dirac action associated to an electron of mass $m$.
 Its action $S_0=\int dx^4\overline{\psi}(i\gamma^{\mu}\partial_{\mu}-m)\psi$ is invariant under constant phase transformations. If one imposes to the global symmetry to become local, that is $\psi\to e^{-i\epsilon}\psi\Rightarrow \psi\to e^{-i\epsilon(x)}\psi
$ then, the action is no longer invariant under the local symmetry and recovering it requires to introduce a gauge field $A_{\mu}$. This transforms as $A_{\mu}\to A_{\mu}+\frac{1}{e}\partial_{\mu}\epsilon(x^{\nu})$ to compensate the local phase transformation. At the same time, one must replace the ordinary derivative by the covariant one $D_{\mu}=\partial_{\mu}+ie A_{\mu}(x^{\nu})$ in order to achieve the invariant new action. At quantum level $e$ is identified with the electric charge. The modified action is
$$
S_1=\int dx^4 \overline{\psi}(i\gamma^{\mu}D_{\mu}-m)\psi
$$
representing the term for the minimal coupling between a gauge boson and an electron in the electrodynamic action. This would be the so called Noether gauging process which is the standard one. Physically, one includes the gauge field dynamics by introducing a self-interaction term for the gauge field. At quantum level this gauge field is interpreted as a gauge boson whose dynamics is described by the electrodynamic action $S_{U(1)}=S_1+\frac{i}{4}\int dx^4 F_{\mu\nu}F^{\mu\nu}.$

\section{Geometry of Sculpting-like processes}
 In \cite{gm} a new mechanism for gauging a theory based on fiber bundle deformation was proposed. The \textit{Sculpting mechanism} process consists in a deformation of bundles $E$ preserving the fiber $F$ and the base $B$, but allowing changes in the bundle structure. We will describe it in three different examples, emphasizing its mathematical requirements in the first two examples and one last example with a realization in M-theory.
\paragraph{Toy Model on a  trivial torus bundle.}
 This first example shows the skeleton of the procedure and its main interest is mathematical.
As a starting point, let us consider  the trivial bundle $\Sigma\times T_F\to \Sigma$ where $T_F$ is a 2-torus over the
 torus $\Sigma=\big({\mathbb R}/{\mathbb Z}\big)^2=\{(x+{\mathbb Z},\,y+{\mathbb Z})\,|\,x, y\in{\mathbb R}  \}.$
 A section of this fiber bundle
 is given by a $C^{\infty}$ map $\xi:\Sigma\to T_F$, which gives rise
 to a pair $(\xi^1,\xi^2)$ of ${\mathbb R}/{\mathbb Z}$-valued maps defined on
 $T_F$. The form $d\xi^j:=\frac{\partial\xi^j}{\partial x}\,dx+\frac{\partial\xi^j}{\partial
  y}\,dy$ is a well-defined closed form on $T_F$, but it is {\em not} an exact
  form, since $\xi^j$ is not a well-defined ${\mathbb R}$-valued function
  on $T_F$. On this space we consider that a physical theory is defined as the invariant functional constructed in terms of the closed one-forms representing the action.  We can consider on $\Sigma$ the metric $dx\otimes dy$ and the orientation defined by the $2$-form $\omega_B= dx\wedge dy$. With respect to this metric a $1$-form on $T_F$ is harmonic iff it is a linear combination $a_1dx+a_2dy$, with $a_1,a_2\in{\mathbb R}$.  On the other hand, the harmonic component, in the Hodge decomposition, of the closed form
  $\gamma=\gamma_1dx+\gamma_2dy$ is
  $\widehat{\gamma}:=\Big(\int_\Sigma\gamma_1(x,y)\,dv\Big) dx+ \Big(\int_\Sigma\gamma_2(x,y)\,dv\Big) dy,$
  where $dv$ is the volume element on $\Sigma$ defined by the metric and the orientation that have been fixed. The exact forms are characterized by periodic real functions on $\Sigma$.\\

Now, we impose a quantization condition over the harmonic forms that will allow us to extract the gauge field connection. This condition has very important consequences at the physical level as we illustrate in the third example. Geometrically it adds a new principal circle bundle $L$ to the torus bundle construction by means of a topological restriction. One considers a section $\xi$ such that the corresponding harmonic forms  $\widehat{d\xi^1}$ and $\widehat{d\xi^2}$, defined above satisfy
 \begin{equation} \label{mreq}
  \int_{\Sigma} \widehat{d\xi^1}\wedge \widehat{d\xi^2}\in{\mathbb Z}.
  \end{equation}
 Thus, the torus $\Sigma$ equipped with the symplectic form $\widehat{d\xi^1}\wedge \widehat{d\xi^2}$ is a quantizable manifold \cite{wood}. That is, there is a Hermitian line bundle $L$ over $\Sigma$ endowed with a connection whose curvature equals $\frac{2\pi}{i}\,\widehat{d\xi^1}\wedge \widehat{d\xi^2}.$
  Hence, the first Chern class $c_1(L)$ is the integer of condition (\ref{mreq}).
 Since  $\widehat{d\xi^j}$ is a  harmonic $1$-form,  $\widehat{d\xi^j}=a^j_1dx+a^j_2dy$, with $a^j_i$ constant, then $\int_{\Sigma} \widehat{d\xi^1}\wedge \widehat{d\xi^2}=a^1_1a^2_2-a^1_2a^2_1.$  Let us denote $\int_{\mathcal{C}_i} \widehat{d\xi}^j=a^j_i$ with $\mathcal{C}_i$ being one-cycles of $T_F$, then, $c_1(L)={\rm det}(a^i_j)$.
Now, we can define the connection $C$ on the principal torus bundle:
The exact forms $(e(\xi^1)$, $e(\xi^2))$ -which are the exact part of $d\xi^i$- allow us to define a $T(orus)$-invariant connection on this trivial $T$-principal bundle. To this end, let consider vectors $u$ of the tangent space at points $x$ of $\Sigma$ and its tangent map $\xi_*$ ($\xi_*$ is the map induced by $\xi$ on the tangent spaces to $\Sigma$). Then, it is sufficient to give the value of the corresponding ${\mathbb R}^2$-valued $1$-form $C$ on $\Sigma\times T_F$ on the vectors which are in the image of $\xi_*$. We define,
\bea C(\xi_*u)=\big(e(\xi^1)(u),\, e(\xi^2)(u)\big).\eea
Obviously, the pullback of $C$ to the basis by the section $\xi$ is the ${\mathbb R}^2$-valued $1$-form $(e(\xi^1),e(\xi^2))$. Since the forms $e(\xi^i)$ are exact, the curvature form of this connection vanishes.
Summarizing, by imposing a topological restriction on the original bundle, a new nontrivial line bundle has emerged and the original trivial torus bundle has changed to a trivial torus bundle with a connection transforming under $U(1)\times U(1)$ (torus group) gauge symmetry.
\paragraph{On a Flat torus bundle.} Let us consider the same toy model on a trivial torus bundle over a torus base manifold and repeat all the previous steps till the construction of the connection. Now we want to obtain a connection on a flat torus bundle $\widetilde{\mathcal{C}}$. To this end, we need to define the monodromy $\rho$ of the bundle. We will do it by adding weight functions $g_i, i=1,2$ to pair of exact one-forms as $(e_i=d\alpha_i),i=1,2$ restricted to satisfy the condition $d(g_id\alpha^i)=0$. Then, it is possible to define a flat connection $\widetilde{\mathcal{C}}$ on the bundle $E$ as follows
$$
\widetilde{\mathcal{C}}(\xi_*u)=(g_1 d\alpha^1,g_2 d\alpha^2),
$$
 The monodromy  of the flat torus bundle is a representation of the fundamental group of base torus on the $Aut(T_f)$, $\rho:\pi_1(T_B)\to Aut(T_F)$.  The gauged theory is then described by an invariant functional on the sections of a flat torus bundle with monodromy contained in the torus group.

\paragraph{Supermembrane theory sculpted on a nontrivial torus bundle.} The Supermembrane theory describes the dynamics of an extended 2+1D M-theory object embedded in 11D target spaces -compact or not-\footnote{The term compactification refers to the wrapping of some of the spatial dimensions.}. Assume a toroidal supermembrane theory formulated in the Light Cone Gauge wrapping a target-space $M_9\times T^2$ where $M_9$ denotes the 9D Minkowski spacetime. The three dimensional worldvolume is assumed to be foliated in $\mathbb{R}\times \Sigma$ where $\mathbb{R}$ parametrizes the proper time $\tau$ and $\Sigma$ denotes a compact Riemann surface  of genus one, with local coordinates $(\sigma^1,\sigma^1), \sigma^i\in [0,2\pi]$. we will consider the embedding as the initial fiber bundle $E$: a trivial bundle with the 11D target-space $M_9\times T^2$ as a fiber and a compact torus as a base. Since it is a worldvolume theory, it is invariant under residual area preserving diffeomorphims (APD), and it has $SL(2,\mathbb{Z})$ as a global symmetry.  We define the standard complex coordinates on the base torus as $z=\sigma^1+i\sigma^2,\overline{z}=\sigma^1-i\sigma^2$. On the base we fix the standard complex structure $J_B^0$ and the \textit{Kh\"aler} form $\omega_B^0=dz \wedge d\overline{z}$. Then, there is an associated \textit{K\"ahler} metric $g_B^0=dz\otimes d\overline{z}$ with determinant $W_0$. We do the same for the torus of the fiber $(J_F,\omega_F,g_F)$ now expressed in terms of the complex coordinate $u=u_1+u_2$ and $\overline{u}$ of $T_F$. The fiber bundle $E$ is a trivial bundle with the 11D target-space $M_9\times T^2$ as a fiber and a compact torus as a base. The embedding maps are $X^M: \Sigma\to T^2\times M_9$, are sections of this bundle $E$ where $X^{M}=(X,X^m,X^+,X^-), m=1,\dots,7$ and $X= X^1+iX^2\in\Gamma(T^2)$, $X^m\in\Gamma(M_9)$, $X^{\pm}$ are the usual light cone coordinates that decouple from the Hamiltonian. The associated Hamiltonian describing the supermembrane is \cite{gmpr}
\begin{equation}
\begin{aligned}
H=&T^{-2/3}\int_{\Sigma}dzd\overline{z}\sqrt{W_0}\left[\frac{1}{2}(\frac{P_m}{\sqrt{W}})^2
+\frac{1}{2}\frac{P\overline{P}}{\sqrt{W}}+\frac{T^2}{4}\{X,X^m\}^2+\frac{T^2}{4}\{\overline{X},X^m\}^2+\frac{T^2}{4}\{X^m,X^n\}^2\right]\\
+&T^{-2/3}\int_{\Sigma}dzd\overline{z}\sqrt{W_0}\left[\frac{T^2}{2}\{X,\overline{X}\}^2-\overline{\Psi}\Gamma_{-}\Gamma_m\{X^m,\Psi\}
-\frac{1}{2}\overline{\Psi}\Gamma_{-}\Gamma\{\overline{X},\Psi\}-\frac{1}{2}\overline{\Psi}\Gamma_{-}\overline{\Gamma}\{X,\Psi\}\right]
\end{aligned}
\end{equation}
where $\int_{\Sigma}dzd\overline{z}\sqrt{W_0}$ in the following denoted as $\int_{\Sigma}$ is the area element, $T\equiv T_{M2}$ is the 11D tension of the supermembrane and has dimensions of $[mass/area]$, $X^m$ are scalars parametrizing the transverse coordinates of the supermembrane, $P_m$ are densities and the canonical momenta associated to the $X^m$, and respectively $P$ that of the field $X$. $\Psi$ are  Majorana fermions, they are scalars on the worldvolume but an $SO(7)$ spinor in the target space. The action is supersymmetric, and the bracket is defined as $\{A,B\}=\frac{\epsilon^{ab}}{\sqrt{W_0}}\partial_a A\partial_b B; a,b=z,\overline{z}$.
The Hamiltonian is subject to the APD group residual constraints (connected to the identity $\phi_1$, but also to the large APD  $\phi_2$)
$$\small\small \phi_1:d(\frac{1}{2}(Pd\overline{X}+\overline{P}dX)+P_mdX^m-\overline{\Psi}\Gamma_{-}d\Psi=0;\quad
\phi_2:\oint_{\mathcal{C}_r}\frac{1}{2}(Pd\overline{X}+\overline{P}dX)+P_mdX^m-\overline{\Psi}\Gamma_{-}d\Psi=0.
$$
 $\mathcal{C}_r, r=1,2$ is the canonical 1-homology basis on $T^2$. Since the target-space is compactified, one wants to impose the \textit{wrapping condition} of the supermembrane around the compact space.
$$\small\oint_{\mathcal{C}_s}dX=2\pi R(l_s+m_s)\tau;\quad l_s,m_s\in\mathbb{Z},$$ where $R,\tau$ are, the radius and the Teichmuller parameter of the fiber torus $T^2$ and $l_s,m_s$ with $r,s=1,2$ the winding numbers that form a constant matrix $\mathbb{W}\in SL(2,\mathbb{Z})$. Now we follow the previous steps: Perform \textit{the Hodge decomposition} of the closed one-forms: $dX=2\pi R(l_s+m_s\tau)\widehat{dX^s}+dA$ with $dA$ the exact one-form, $d\widehat{X}= d\widehat{X}^1+id\widehat{X}^2$ the harmonic one-form. Since the maps satisfy the winding condition, there exist global symmetries in the action associated to the invariance under the change of the basis of harmonic functions. The closed one-forms, in particular the exact one-forms (that will be now the dynamical degrees of freedom of the gauged theory), are also going to transform under the group of symplectomorphisms of the base.
Next step is to impose the \textit{nonvanishing central charge condition} (the prequantizable condition)
$$
\int_{\Sigma} dX\wedge \overline{dX}=n,$$ but now restricted to $n\in\mathbb{Z}\setminus\{0\}$ to guarantee the discreteness of the spectrum -as introduced initially in \cite{mrt} and fully proved in \cite{bgmr}-. This condition implies the existence of a nontrivial line bundle $L$, with first Chern class $c_1=n$ \cite{monopole}.

Now, we introduce a \textit{deformation of the complex structure of the base manifold} by imposing a link between the properties of the base and the compact part of the fiber. By fixing an integer basis of the torus homology $(\alpha,\beta)$ one can define the holomorphic one-forms:
$\oint_{\beta}h=\tau;\quad \oint_{\alpha}h=1.$ Then, we impose the identification of the base torus and that of the fiber by making the pullback of the symplectic form of the fiber in the base manifold through some holomorphic maps $\widehat{dX}^i\in \widehat{d\xi}^i$ which satisfy $h=\widehat{dX^1}+i\widehat{dX^2}$. By means of this identification there appear new induced symplectic forms of the base $\omega_{B}=\omega_{rs}\widehat{dX^r}\wedge\widehat{dX^s}$ and a new induced metric on the base $g_{ab}^B=\widehat{\partial_a X^r}\wedge\widehat{\partial_b X^r}\widetilde{g}_{rs}^F$ (see \cite{br} for more details), and most importantly, the symplectic bracket can now be expressed in terms of either the fiber or the base
$\{\phi,\varphi\}=\omega_{B}^{ab}\partial_a\phi\partial_b\varphi=\omega_F^{rs}D_r\phi D_s\varphi.$
This last condition is very relevant from the physical point of view since it imposes a relation between the global symmetries of the base and those of the fiber,\footnote{in String theory, it implies a relation between p-branes and target-space.}. Geometrically, it corresponds to minimal embeddings. This construction was developed at \cite{br}.\\
By restricting the harmonic forms to a subset of holomorphic harmonic forms we induce a change in the complex \textit{K\"ahler} manifold $J_B^0\to J_B^1$. By Kodaira-Spencer, this change of the complex structure is associated with an element of the first cohomology group of the sheaf of germs of holomorphic vector fields on $\Sigma$. It is very relevant that the diffeomorphism class preserving the new complex structure may change to an inequivalent class $Diff(\Sigma_0)\to Diff(\Sigma_1)$. Indeed, the new area determinant is now defined in terms of the pullback of the metric in the fiber onto the base $\sqrt{W}=\frac{1}{2}\widehat{\partial_a X}\widehat{\partial_b \overline{X}}\epsilon^{ab}$.\\
Physically, the theory is invariant under the symplectomorphism transformations \cite{mor} and we want the connection of the base to transform under it. Consequently it is  natural to formulate it terms of a principal bundle with this structure group such that the torus bundle will be associated to it and it will become a \textit{symplectic torus bundle}. Associated to the symplectic form, it can be always defined a nontrivial  symplectic principal bundle whose fiber is the symplectomorphism group. Let us consider a change in the map between the line bundle and the associated torus bundle. Initially, the map satisfies, $\gamma:G_1=Symp\to \{Id\}$ connecting the line bundle and the principal torus bundle. Now it is  modified into $ \widetilde{\gamma}:G_1\to Diff(T^2)$, through the definition of the new transition functions $\widetilde{\varphi_{ij}}$ between two charts $U_i$ and the map $\gamma$ as follows, $\widetilde{\varphi_{ij}}=\widetilde{\gamma}(\varphi_{ij}(x),)\quad \forall x\in (U_i\cap U_j)\subset M$, defined in terms of the old ones $\varphi_{ij}$ .\\

The symplectic torus bundle connection $C=(e(\xi^1),e(\xi^2))$ is inherited from the symplectic connection defined on the symplectic line bundle. The pullback of the connection defines a gauge field on the base manifold whose symplectic covariant derivative is defined as
$\mathcal{D}\bullet=D\bullet+\{A,\bullet\}$ where $D$ is a new rotated covariant derivative defined as $D=e^a\partial_a$ with  $e^a=e_1^a+e_2^a$, given by the zwei-bein $e_r^a=-\omega_B^{ab}(\widehat{\partial_b X^s})\widetilde{g_{rs}}, r=1,2$.
The symplectic curvature defined on the base manifold is
$\mathcal{F}=D\overline{A}-\overline{D}A+\{A,\overline{A}\}.$ It transforms as a connection under the symplectomorphism transformation $\delta_{\epsilon}A=\mathcal{D}\epsilon$. See \cite{mr} for a detailed analysis. \\
In \cite{khan} it was stated that for any fibre bundle $\chi: F \to E \to B$ with structure group $G$, the action of G on F produces a $\pi_0(G)$-action on the homology and cohomology of $F$. For $F$ being the 2-torus with structure group $G=Symp(T^2)$ it is shown that $\pi_0(G)\approx SL(2,\mathbb{Z})$. The $\pi_0(G)$-action has a natural action on $H_1(T^2)$ of $SL(2,\mathbb{Z})$ on $Z^2$\footnote{$Z^2$ represents a pair of integers $(a,b)$ characterizing the $H_1(T^2)$ charges. In \cite{gmpr} these charges were interpreted as the quantized Kaluza Klein momenta  of the compactified supermembrane. }. It exists then a representation $\rho:\pi_1(\Sigma)\to\pi_0(G)$, therefore there is a natural, bijective correspondence between the equivalence classes of symplectic torus bundles over $B$ inducing the module structure
 $\mathbb{Z}_{\rho}$ on $H_1(T^2)$ and the elements of the second cohomology group of $B$, $H^2(B,\mathbb{Z}_{\rho})$. The local coefficients $\mathbb{Z_{\rho}}$ run only over the integers allowed by the representation $\rho(n,m)$, see \cite{khan} for more details. We call this representation  $\rho$ \textit{monodromy}. In particular this construction holds when the base manifold is the torus like in the case we are considering. The cohomology  of the bundle is in general nontrivial and it depends on the coinvariant class inside each of the conjugate classes of the symplectic torus bundle with monodromy in $SL(2,\mathbb{Z})$ \cite{gmmpr},\cite{gmpr}.\\
The symplectic connection defined on the base manifold transforms with the monodromy, and it does as $dA\to dA e^{i\varphi_{\rho}}$ where $\varphi_{\rho}$ is a discrete monodromy phase given by $\varphi_{\rho}=\frac{c\tau+d}{\vert c\tau+d\vert}$ for a given modulus $\tau$, where the monodromy matrix is $\rho=\begin{pmatrix}a&b\\c&d\end{pmatrix}\in SL(2,\mathbb{Z})$. The covariant derivative transforms in terms of a matrix
$D_r\bullet=(2\pi R_rl_r)\theta_{lr}\frac{\epsilon^{ab}}{\sqrt{W(\sigma)}}\widehat{\partial_a X^l(\sigma)}\partial_b\bullet$. Due to fixing of the base and the fiber $\theta\in SL(2,\mathbb{Z})$ encodes the discrete residual symmetries determined by the monodromy (see \cite{gmpr} for more details). The resulting Hamiltonian is invariant under the monodromy of the bundle, and represents the sculpted supermembrane:
\begin{equation}
\begin{aligned}
\small
H=&T^{2/3}\int_{\Sigma}\left[\frac{1}{2}(\frac{P_m}{\sqrt{W}})^2
+\frac{1}{2}\frac{P\overline{P}}{\sqrt{W}}+\frac{T^2}{4}\{X^m,X^n\}^2+\frac{T^2}{2}(\mathcal{D}X^m)(\mathcal{\overline{D}}X^m)
+\frac{T^2}{4}(\mathcal{F}\overline{\mathcal{F}})\right]
\\+&(n^2Area_{T^2}^2)+T^{2/3}\int_{\Sigma}\left[-\overline{\Psi}\Gamma_{-}\Gamma_m\{X^m,\Psi\}
-\frac{1}{2}\overline{\Psi}\Gamma_{-}\overline{\Gamma}\mathcal{D}\Psi\}-\frac{1}{2}\overline{\Psi}\Gamma_{-}\Gamma\overline{\mathcal{D}}\Psi\}\right]
\\+&\int_{\Sigma}\sqrt{W}\Lambda\left[\frac{1}{2}\overline{\mathcal{D}}(\frac{P}{\sqrt{W}})+\frac{1}{2}\mathcal{D}(\frac{\overline{P}}{\sqrt{W}})+\{X^m,\frac{P_m}{\sqrt{W}}\}-\Lambda\{\overline{\Psi}\Gamma_{-},\Psi\}\right]
\end{aligned}
\end{equation}
where $\Lambda$ is a Lagrange multiplier.
The gauged theory correspond to a supermembrane formulated on a nontrivial symplectic torus bundle with monodromy in $SL(2,\mathbb{Z})$, discrete spectrum (unlike the undeformed theory), global symmetries restricted by the monodromy and new degree of freedom: a gauge field which transform with the symplectomorphisms group of the fiber. The bundle is in the nontrivial cohomology class $H^2(\Sigma,\mathbb{Z}_{\rho})$ \cite{gmpr}.
\section{Discussion}
We have shown the geometrical construction of the sculpting-like gauging methods on three different examples on torus bundles. In all of them a gauge connection (with or without monodromy) was extracted as a dynamical  field of a modified theory by mainly performing a Hodge decomposition and imposing several restrictions on the fields of the theory that led to an invariant new action on a modified bundle. In this way a topological change at the homotopy-type on the bundle is induced such that the homotopy-type of the base and fiber is preserved. These Sculpting-like methods are quite natural in theories containing extended objects like branes in String/ M-theory  but also for theories over compact spaces. More generally it applies to actions with closed one-forms. A natural generalization is to apply to compact p-branes with p larger than 2, and a Nambu-Goto like actions, by splitting the Nambu-Poisson brackets  into nested Lie brackets \cite{m5}. It can also be done for DBI actions by expanding the determinant and working out the part of the action associated to the metric along the lines explained in \cite{gm}. In this fashion, one could explore, as an example -since the local description is already done in \cite{mor2}\cite{br}-, the toroidal $D4$ brane embedded in $M_6\times T^4$ with the LCG canonical hamiltonian and subject to a monopole condition.
\section{Acknowledgements} The author want to strongly thank prof. Andr\'{e}s  Vi\~{n}a for discussions over many geometrical aspects and collaboration in the formalization of these results and to the reviewer for his/her careful reading and help to improve the manuscript. The author would also like to thank the IFT for a kind invitation and for financial support through projects S2009/ESP-1473 (HEPHACOS), FPA2012-32828 and to the Univ. of Oviedo through the project  SV-PA-13-ECOEMP-30.

\end{document}